\documentclass[twocolumn]{aastex63}
\usepackage[utf8]{inputenc}
\usepackage {graphicx}

\newcommand\teff{\mbox{$T_{\rm eff}$}}
\newcommand\lbol{\mbox{$L_{\rm bol}$}}

\begin{document}

\title{Spectroscopic Characterization of WD J000801.25-350450 and its Two Co-Moving Companions}

\correspondingauthor{Peter A. Jałowiczor}

\author[0000-0002-4175-295X]{Peter A. Jałowiczor}
\affil{Backyard Worlds: Planet 9}

\author[0000-0003-2235-761X]{Thomas P. Bickle}
\affil {School of Physical Sciences, The Open University, Milton Keynes, MK7 6AA, UK}

\author[0000-0003-4269-260X]{J. Davy Kirkpatrick}
\affil {IPAC, Mail Code 100-22, Caltech, 1200 E. California Blvd., Pasadena, CA91125, USA}

\author[0000-0003-2478-0120]{Sarah L. Casewell}
\affil{School of Physics and Astronomy, University of Leicester, Leicester, LE1 7RH, UK}

\author [0000-0002-6428-4378] {Nicola Gentile Fusillo}
\affil{Department of Physics, Università degli Studi di Trieste, Trieste, 34127, Italy }
\affil{INAF-Osservatorio Astronomico di Trieste, Via G.B. Tiepolo 11, I-34143 Trieste, Italy}

\author[0000-0002-6294-5937]{Adam C. Schneider}
\affil{United States Naval Observatory, Flagstaff Station, 10391 West Naval Observatory Rd., Flagstaff, AZ 86005, USA}

\author[0000-0002-2592-9612]{Jonathan Gagn\'e}
\affil{Plan\'etarium de Montr\'eal, Espace pour la Vie, 4801 ave. Pierre-de Coubertin, Montr\'eal, QC H1V~3V4, Canada}
\affil{Trottier Institute for Research on Exoplanets, Universit\'e de Montr\'eal, 2900 Boulevard \'Edouard-Montpetit Montr\'eal, QC H3T~1J4, Canada}

\author[0000-0001-6251-0573]{Jacqueline K. Faherty}
\affil{Department of Astrophysics, American Museum of Natural History, Central Park West at 79th Street, NY 10024, USA}

\author[0000-0002-1125-7384]{Aaron M. Meisner}
\affil{NSF National Optical-Infrared Astronomy Research Laboratory, 950 N. Cherry Ave., Tucson, AZ 85719, USA}

\author[0000-0002-2387-5489]{Marc J. Kuchner}
\affil{Exoplanets and Stellar Astrophysics Laboratory, NASA Goddard Space Flight Center, 8800 Greenbelt Road, Greenbelt, MD 20771, USA}

\author[0000-0002-6523-9536]{Adam J.\ Burgasser}
\affiliation{Department of Astronomy \& Astrophysics, UC San Diego, La Jolla, CA, USA}

\author[0000-0003-4083-9962]{Austin Rothermich}
\affiliation{Department of Astrophysics, American Museum of Natural History, Central Park West at 79th Street, NY 10024, USA} 

\author[0009-0002-3936-8059]{Alexia Bravo}
\affiliation{United States Naval Observatory, Flagstaff Station, 10391 West Naval Observatory Rd., Flagstaff, AZ 86005, USA}

\author[0000-0001-8343-0820]{Michiharu Hyogo}
\affil{Backyard Worlds: Planet 9}

\author[0000-0001-9482-7794] {Mark Popinchalk}
\affiliation{Department of Astrophysics, American Museum of Natural History, Central Park West at 79th Street, NY 10024, USA}

\author [0000-0002-3316-7240]{Alex J. Brown}
\affiliation{Hamburger Sternwarte, University of Hamburg, Gojenbergsweg 112, 21029 Hamburg, Germany}

\author [0000-0002-6153-7173] {Alberto Rebassa-Mansergas}
\affiliation{Departament de F\'isica, Universitat Polit\`ecnica de Catalunya, c/Esteve Terrades 5, 08860 Castelldefels, Spain,  Institute for Space Studies of Catalonia, c/Gran Capit\`a 2--4, Edif. Nexus 104, 08034 Barcelona, Spain}

\author [0009-0002-8578-0765] {Raquel Murillo-Ojeda}
\affiliation{Centro de Astrobiolog\'ia (CAB), CSIC-INTA, Camino Bajo del Castillo s/n, campus ESAC, 28692, Villanueva de la Ca\~nada, Madrid, Spain}

\author{The Backyard Worlds: Planet 9 Collaboration}
\affil{backyardworlds.org}

\newcommand{\masslmssun}{$0.077^{+0.005}_{-0.017}$\,$M_{\sun}$}
\newcommand{\masslmsmjup}{$80.2^{+0.5}_{-1.8}$\,$M_{\rm Jup}$}
\newcommand{\totagewd}{5.5$^{+4.3}_{-1.9}$\,Gyr}
\newcommand{\teffwd}{5903$\pm$22\,K}
\newcommand{\loggwd}{$\log{g}$ = 8.040$\pm$0.008}
\newcommand{\initmass}{$1.5 \pm 0.4$\,$M_{\sun}$}

\begin{abstract}
We present new spectroscopic data for Gaia DR3 2309499817384726016 (WD0008-350A) and its two wide, co-moving, low-mass companions.
We confirm the white dwarf is a hydrogen rich DA, with T$_{\rm eff}$=6200$\pm$90~K and a mass of 0.63$\pm$0.03~M${\odot}$, close to that of the average white dwarf.

Near-infrared spectra of the two stellar companions to WD0008-350A reveal that the inner companion is an M dwarf, exhibiting a spectral type of M8. Furthermore, the outer companion is identified as a possible M6 + M9 binary. This paper examines the evidence which suggests the system may be quadruple. 

\end{abstract}


\section{Introduction}

A significant proportion of the stars in the Milky Way are found in multiple systems. The reasons for the different hierarchies are crucial to understanding the stellar formation process, and multiple systems containing white dwarfs are important to understand how these systems evolve. White dwarfs in multiple stellar systems are essential for investigating how stellar interactions with companion stars influence stellar evolution \citep{10.1093/mnras/stac3675}. Studying hierarchical systems helps refine stellar evolution models in multiple systems. This includes processes such as the evolution of white dwarf binaries during any mass-transfer phase \citep{2004ApJ...601.1058I}, tidal interactions, and common-envelope phases \citep{Iben_1993} as well as the initial formation of the system \citep{Mason_2018}.


Backyard Worlds: Planet 9 (BYW: P9) \citep{kuchner2017} uses citizen science to search images from the {\it Wide-field Infrared Survey Explorer} mission ({\it{WISE}}; \citealt{wright2010}) for moving objects: brown dwarfs and stars with high proper motions, and new planets in the outer solar system \citep[e.g.][]{Batygin2016AJ}. The project identified 112 objects in the 20 pc full-sky census of \cite{Kirkpatrick2024}. Since its inception, the BYW:P9 project has expanded its scope, resulting in the Backyard Worlds: Cool Neighbours project. This uses machine learning to discover Y dwarf candidates.  BYW: P9 has also seen success with the discovery of exotic objects such as extreme T-type subdwarfs \citep{schneider2020, Meisner_2021}, low-mass companions to other stars \citep{rothermich2024, Faherty_2021}, and previous discoveries in the white dwarf field \citep{debes2019, jalowiczor2021, Bickle}. 

In this paper, we discuss Gaia DR3 2309499817384726016  or WDJ000801.25-350450.2 (WD0008-350A), comprised of a WD primary, in a higher-order hierarchy with 2 (or 3) low-mass stellar companions.  We independently identified this system as part of the BYW: P9 science project, hosted on the Zooniverse platform. We have characterized this newly discovered system with spectroscopy of each component. 
The system is described in Section 2. 
The observations of the system are described in Section 3, and the analysis of the new data is described in Section 4. The system, in the context of other known WD multiples, is discussed in Section 5.

\section{The WDJ000801.25-350450.2 system}

\begin{figure}[ht]
    \centering
    \includegraphics[scale=0.15] {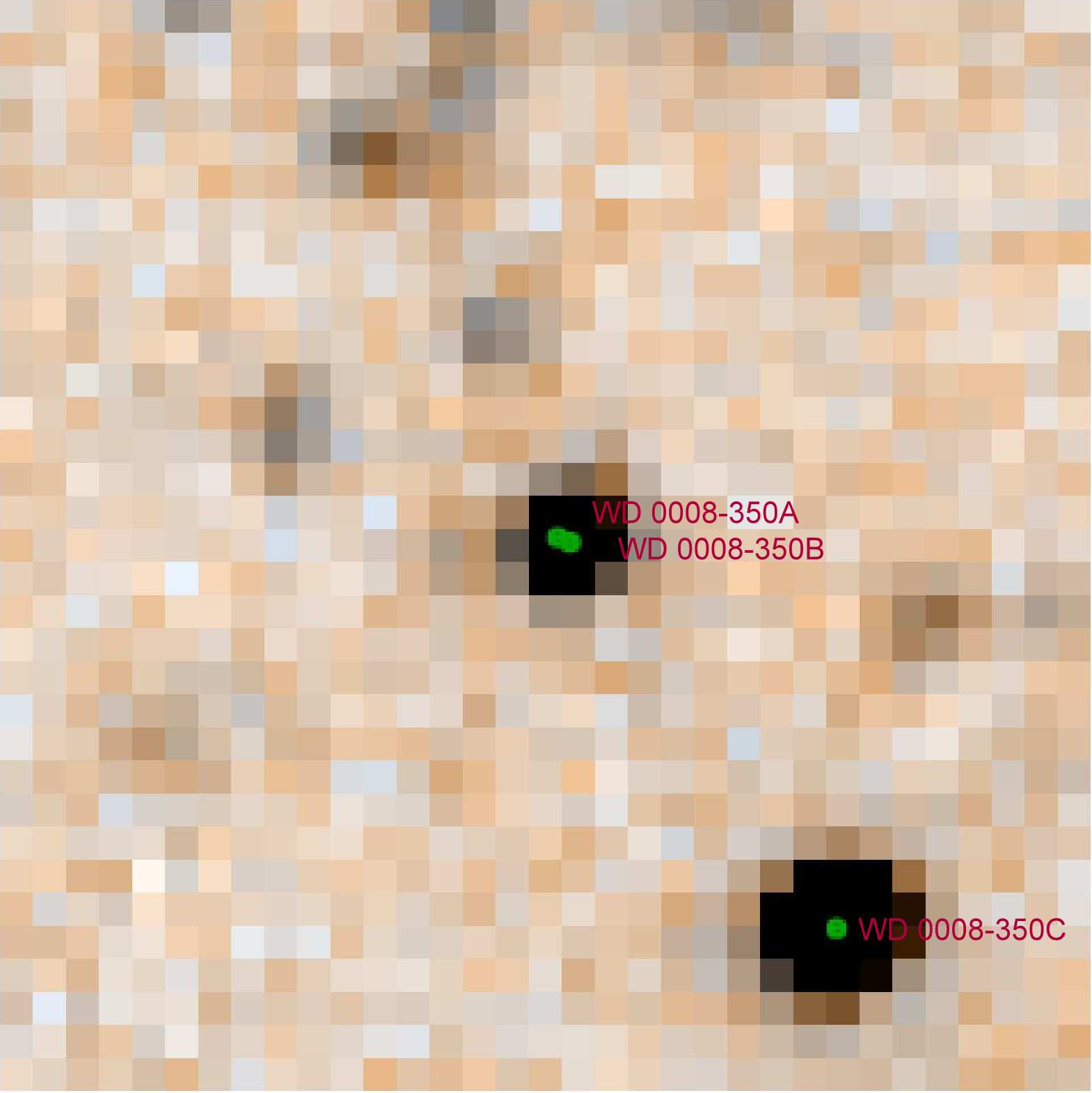}
    \caption{The "Family Portrait", a pseudo-colour composite image of this system. This was created from the Wide-field Infrared Survey Explorer (WISE) instrument bands. Blue represents the W1 band. Red represents the W2 band. Green represents the average of the W1 and W2 bands.  The green dots are the Gaia Overlay; they indicate that a celestial object has already been catalogued in the Gaia database.     
    The field of view is 90 arcseconds $\times$ 90 arcseconds.
    }

\label{Family Portrait}
\end{figure}

Fig.\ref {Family Portrait} was constructed using the \texttt{WiseView} tool \citep{Wiseview} from data acquired by the Wide-field Infrared Survey Explorer (WISE) mission. The image is a composite of unWISE coadds \citep{tr_coadds} from two specific mid-infrared filter bands. These are the WISE W1 (centered at 3.4 ($\mu$m)) and W2 bands (centered at 4.6 ($\mu$m)). This combination helps visualise objects with different infrared properties, since the W1 and W2 bands are centered on different infrared wavelengths.

\cite{nicola21} first identified  WD0008-350A with a probability of being a white dwarf of 0.78,  
T$_{\rm eff}$=4673$\pm$180~K and log g=6.975$\pm$0.233 and a mass of 0.18$\pm$0.06 M$_{\odot}$. \cite{vincent2024} subsequently used the $Gaia$ XP spectra to determine that WD0008-350A is a Hydrogen-rich DA white dwarf with T$_{\rm eff}$=5378$\pm$108~K and log g= 7.396$\pm$0.063 with a mass of 0.292$\pm$0.033 M$_{\odot}$. The masses determined by both \cite{nicola21} and \cite{vincent2024} are very low for a white dwarf, well below the average of $\sim$0.6~M$_{\odot}$ \citep{kepler2016whitedwarfmassdistribution}. The presence of the close companion Gaia DR3 2309499813089120512 at 1\farcs1 away presents a potential source of photometric contamination that could have influenced the $Gaia$ photometry. The $G_{BP}$ magnitude is faint, and the contaminating target is an M dwarf, making the total flux appear brighter at longer wavelengths, suggesting a cooler, but physically larger white dwarf. 

WDJ000801.25-350450.2 was also identified as part of a multiple system by \cite{Tokovinin_2022}, who listed it in their Gaia triple system compilation. They list the white dwarf as component B, although they do not acknowledge it as a white dwarf, and present two suggested companions, Gaia DR3 2309499813089120512 (C) and Gaia DR3 2309499778729380480 (A), separated from the primary by 1\farcs1 and 39\farcs48, respectively.	They determined the masses of the three components to be 0.13, 0.10 and 0.14 M$_{\odot}$ from their absolute magnitudes, assuming all components are on the main sequence.  

\cite{Kervella} also identified this system using proper motion anomaly as a primary selection criterion for targeted direct imaging campaigns. Candidates exhibiting a significant proper motion anomaly were identified as promising candidates for hosting a companion.  They determined the respective masses in this system to be 0.11, 0.11 and 0.14 M$_{\odot}$.


\section{Observations}

\subsection{VLT/X-Shooter}

WD 0008-350A was observed on UT 2024 September 02 with the X-Shooter spectrograph \citep{Vernet2011} mounted on UT2 of ESO's Very Large Telescope (VLT) located in Paranal, Chile as part of program 113.26HT.001 (PI: Casewell). We obtained spectra in the UVB (0.30$-$0.56 \micron) and VIS (0.56$-$1.01 \micron) arms using slit widths of 1\arcsec and 0.9\arcsec, respectively, on the night of 2 September 2024 and using an exposure time of 660~s in stare mode. We reduced the data using the standard \textsc{esoreflex} pipeline to perform flat fielding, bias correction, and apply the wavelength correction.

\subsection{IRTF/SpeX}

WD 0008-350B and WD 0008-350C  were observed on UT 2024 August 13 using the SpeX spectrograph \citep{Rayner2003} on NASA's Infrared Telescope Facility (IRTF) telescope. SpeX was operated in prism mode using the $0\farcs8\times15\arcsec$  slit, providing an average resolution of $\sim$200 over the 0.8$-$2.5 $\micron$ wavelength range. The objects were observed using an AB nod pattern along the slit. For the inner companion, we obtained eight 150 s exposures, and for the outer companion, we obtained eight 90 s exposures. Due to their proximity on the sky, the two companions were observed consecutively, at an average airmass of 1.76. For telluric correction, the A0 star HD 225200 was observed immediately after the targets, followed by internal flat-field and arc lamp exposures for calibration. Data were reduced using the \texttt{Spextool} package \citep{Cushing2004} with telluric correction and flux calibration of the A0 star performed using procedures described in \cite{Vacca2003}.

\begin{deluxetable*}{lrrcc}
\tablecaption{System Properties}
\label{table1}
\tablehead{
\colhead{Parameter} & \colhead{WD 0008-350A} & \colhead{WD 0008-350B} & \colhead{WD 0008-350C} & \colhead{Ref.}}
\startdata
Mass ($M_{\odot}$) & 0.63$\pm$0.03 & 0.08 & 0.10 & 0 \\
R.A. (degrees) &  2.0057513 & 2.0054037 & 1.997907 & 1\\
Dec. (degrees) & -35.080540  &-35.0806593 & -35.0895739 & 1\\
\teff\ & {6200$\pm$90\,K} & - & - & 2\\
\teff\ & - & 2572$\pm$190\,K & 2660$\pm$184\,K & 3\\
log g & {8.08$\pm$0.6} & - & - & 0 \\
$\text{RUWE}$ & 1.000  & 1.029 & 1.205 & 1\\
$\varpi$ (mas) & 11.00$\pm$0.20 & 11.10$\pm$0.6 & 10.80$\pm$0.2 & 1\\
$\mu$$_{\alpha}$ (mas yr$^{-1}$) & 92.92$\pm$0.14 & 90.7$\pm$0.4 & 91.55$\pm$0.17 & 1\\
$\mu$$_{\delta}$ (mas yr$^{-1}$) &-19.61$\pm$0.63 & -15.7$\pm$0.8 & -17.80$\pm$0.23 & 1\\
$G$ (mag) & 18.73$\pm$0.003 & 20.21$\pm$0.008 & 18.32$\pm$0.004 & 1\\
$G_{\rm Bp}$ (mag) & 19.02$\pm$0.030 & - & 20.73$\pm$0.123 & 1\\
$G_{\rm Rp}$ (mag) & 17.83$\pm$0.026 & - & 16.82$\pm$0.009 & 1\\
$J_{\rm}$ (mag) & - & 15.76$\pm$0.07  & 14.42$\pm$0.03 & 4\\
$H_{\rm}$ (mag) & - & 15.02$\pm$0.10  & 13.75$\pm$0.02 & 4\\
$K_{\rm}$ (mag) & - & 14.62$\pm$0.10  & 13.39$\pm$0.03 & 4\\
$W1_{\rm}$ (mag) & - & 14.30$\pm$0.03 & 13.16$\pm$0.02 & 5\\
$W2_{\rm}$ (mag) & - & 14.10$\pm$0.04 & 12.93$\pm$0.03 & 5\\
\enddata
\tablerefs{(0) This work (1) Gaia DR3 \citep {refId03} (2)  \cite {Tremblayetal13} (3) \cite {sanghi2023hawaiiinfraredparallaxprogram}
(4) \cite {article} (5) \cite {Wright_2010} } 

\end{deluxetable*}

\section{Analysis}
\subsection{WD 0008-350, Primary}
To obtain a set of spectroscopic stellar parameters for the white dwarf, we fit the continuum-normalised Balmer line profiles in the X-Shooter UVB and VIS spectra with the grid of 3D pure-hydrogen (DA) local thermal equilibrium (LTE) atmosphere models described in \cite{Tremblayetal13} and \cite{Tremblay2015a}. 
A complete description of models' input physics is available on P.-E. Tremblay's Source Model Data webpage\footnote{\url{https://warwick.ac.uk/fac/sci/physics/research/astro/people/tremblay/modelgrids/}}. 

\begin{figure}[ht]
    \centering
    \includegraphics[scale=0.45] {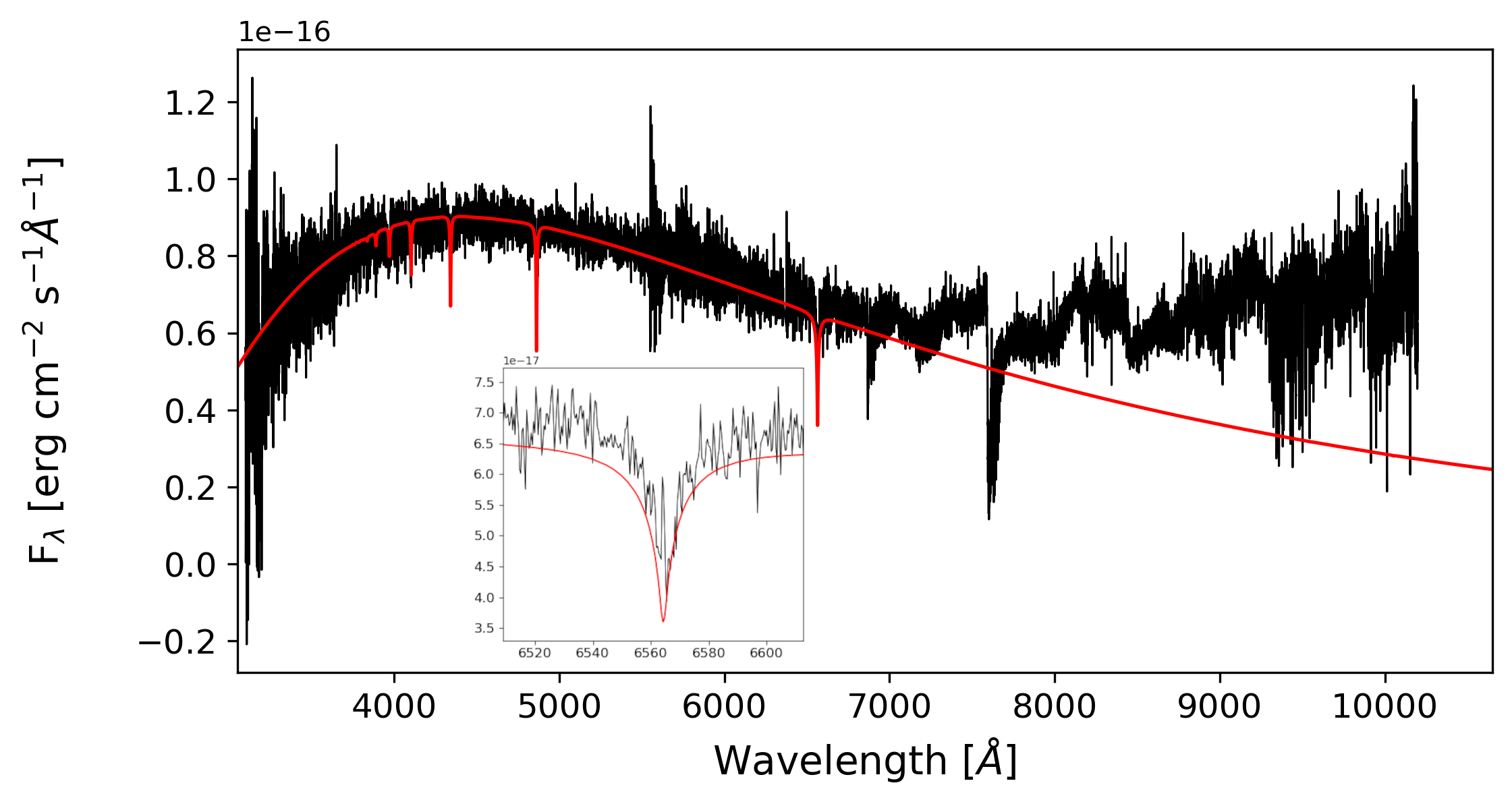}
    \caption{X-Shooter spectrum of WD 0008-350A. The best-fit model to the white dwarf is shown in red overlay. The spectrum is blended with WD 0008-350B, evidenced by the flux excess past $\sim$7000~\AA. The sub-figure provides a magnified view of the H-$\alpha$ line.
    }
    
    \label{fig:enter-label}
\end{figure}

From our best-fitting model, we determine  T$_{\rm eff}$=6200$\pm$90~K and log g=8.08$\pm$0.6. These values are presented in Table \ref{table1}. Fig. \ref{fig:enter-label} shows the spectral fitting to the white dwarf (red overlay), noting that the white dwarf's spectrum is partially blended with that of the inner companion (B) beyond about 7500\AA.

Although the effective temperature is broadly consistent with the previous determinations of \cite{nicola21} and \cite{vincent2024}, the gravity determination has increased significantly with the X-Shooter spectra. This is likely due to the much higher signal-to-noise ratio of the spectrum as well as the broader wavelength coverage than that provided by $Gaia$. However, XP spectra offer a more enhanced spectral classification along with more precise determinations of the effective temperature and surface gravity.

We used \texttt{wdwarfdate} \citep{Kiman_2022} which, when provided with effective temperature, surface gravity and their respective uncertainties, uses the MIST isochrones \citep{Choi_2016, Dotter_2016}, the \cite{cummings19} initial-final mass relation and the \cite{Bedard2020} cooling models to estimate the cooling age, total age, initial main-sequence mass and the current mass of a white dwarf. Using the DA model, which incorporates the DA mass-radius ($M$-$R$) relation with thick hydrogen layers and carbon oxygen cores of \citet{Bedard2020}\footnote{\url{https://www.astro.umontreal.ca/~bergeron/CoolingModels/}}, we determine a total age for WDJ0008-350 of $4.67^{+3.58}_{-0.85}$ Gyr, with a cooling age of $2.45_{-0.39}^{+0.24}$ Gyr. We estimate a progenitor mass of $1.65 _{-0.44}^{+0.56}$M$_{\odot}$ and a current mass of $0.63\pm0.03$ M$_{\odot}$.
For the purposes of this analysis, we assume solar metallicity and $v/v_{\text{crit}}=0$, where $v/v_{crit}$ represents the object's velocity as a fraction of its critical (or break-up) velocity.


The absence of a detectable H-beta (H$\beta$) emission feature in the spectrum (Fig. \ref{fig:enter-label}) strongly suggests that the apparent feature at the H-alpha (H$\alpha$) wavelength is not a true emission line.



In Table \ref{table1}, the RUWE for WD 0008-350A=1.000. This indicates the observed star's motion is well-modeled by the single-star solution \citep {Pearce2022RUWE}.

\subsection{WD 0008-350B, Inner Companion}
\begin{figure}[htbp]
    \centering
    \includegraphics[scale=0.42]{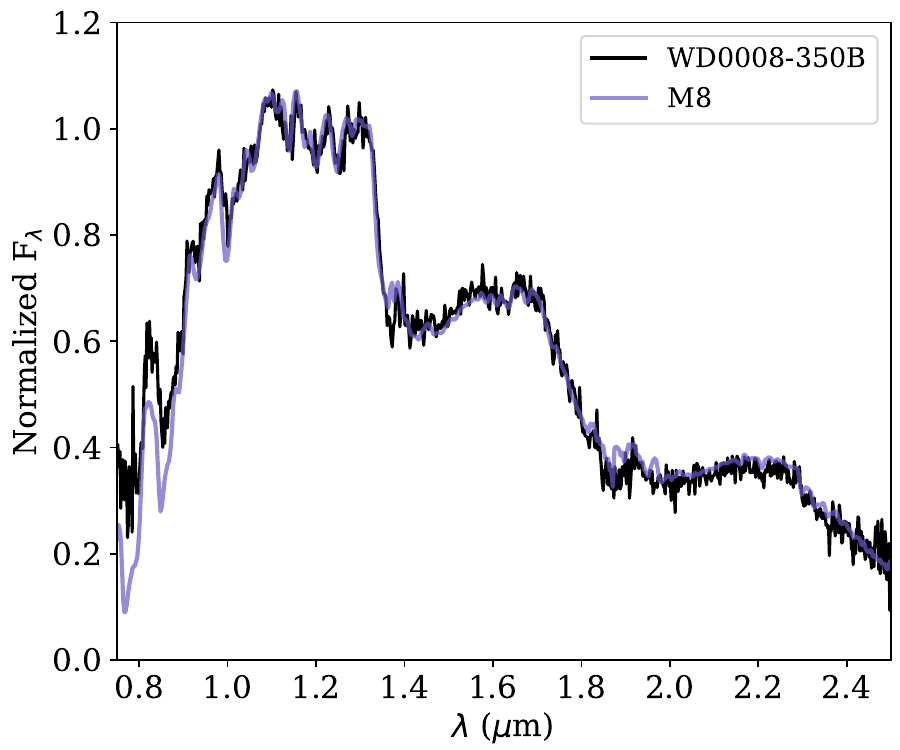}
    \caption{IRTF/SpeX spectrum of the inner companion (black) compared to the M8 spectral standard, VB 10 (blue) from \cite{kirkpatrick2010}. All spectra are normalized between 1.27 and 1.29 $\micron$}
    \label{fig:companion_spectra}
  \end{figure}

To determine the spectral type of the inner companion, Gaia 2309499813089120512 (WD 0008-350B), its SpeX spectrum was compared to M dwarf near-infrared spectral standards from \cite{kirkpatrick2010}. All spectra were normalized between 1.27$-$1.29 $\micron$ and visually compared to determine the best overall fit. The best fit was determined to be the M8 standard (VB 10), which is plotted against the spectrum of the WD 0008-350B in Figure \ref{fig:companion_spectra}. The fit is excellent over the 0.9$-$2.4 $\micron$ range, though WD 0008-350B displays some flux excess at wavelengths $<0.9$ $\micron$, potentially due to contamination from the white dwarf.

We derived the effective temperature (\teff) using empirical relationships for bolometric luminosity (\lbol) and \teff\ as presented in \cite{sanghi2023hawaiiinfraredparallaxprogram}. This gives an effective temperature of 2572$\pm$190 K for the inner companion. 

\cite{2013ApJS..208....9P} presents a collection of empirical relationships and tables that link the effective temperature \teff to the star's spectral type and colour. The stellar mass is subsequently determined by positioning the star on a Hertzsprung-Russell (H-R) diagram and comparing its location, as indicated by its luminosity and temperature, to theoretical evolutionary tracks from stellar models. 

Utilizing stellar evolutionary tracks in \cite{2013ApJS..208....9P}, our analysis resulted in a mass of 0.08 M$_{\odot}$ for WD 0008-350B. This value was obtained by identifying where the star's effective temperature (\teff) intersected with the appropriate evolutionary track on a model Hertzsprung-Russell (H-R) diagram.



In Table \ref{table1}, the RUWE for WD 0008-350B is 1.029; this also indicates a good fit to the single-star solution \citep {Pearce2022RUWE}.

\subsection{WD 0008-350C, Outer Companion}
The outer companion was identified by \cite{Reyle} as an ultracool dwarf candidate based on its position on Gaia DR2 \citep{Gaiadr2} and 2MASS \citep{2mass} color-magnitude and color-color diagrams. They determined a photometric spectral type of M7.  

We compared the SpeX spectrum of Gaia DR3 2309499778729380480 (WD 0008-350C) to near-infrared M and L dwarf spectral standards from \cite{kirkpatrick2010}.  The single best-fitting standard for the C component was the M6 standard LHS 1375. 

For a single template, the M6 standard exhibited an overall agreement with a $\chi^2$ of 30, with the M5 and M7 standards giving $\chi^2$ values of 126 and 43, respectively. We also assessed binary template configurations to attain the highest degree of agreement with the empirical spectra using the methods in \cite{Bravo_2025}.

We applied near-infrared M dwarf standards from \cite{kirkpatrick2010}, combined with 2MASS near-infrared photometry and $Gaia$ DR3 parallax measurements, to create template binary spectra. The best-fitting binary template was M6 + M9, where the M9 standard is LHS 2924.  The M6 + M9 solution is a significantly better fit ($\chi_{\nu}^{2}$ = 9.3) than the single-spectrum M6 solution ($\chi_{\nu}^{2}$ = 30.1). Consequently, the high value of $\chi_{\nu}^{2}$ = 30.1 demonstrates that the single-spectrum M6 solution is statistically improbable and should be rejected in favor of a more robust model. The binary spectral templates resulted in a significantly lower $\chi_{\nu}^{2}$ value, which indicates a more robust and physically plausible fit. For both the single and binary fits, we exclude wavelength regions between 1.35 and 1.45 $\mu$m as well as between 1.81 and 1.95 $\mu$m because these regions can be heavily impacted by telluric absorption.

\begin{figure}[htbp!]
    \centering

    \includegraphics[scale=0.5]{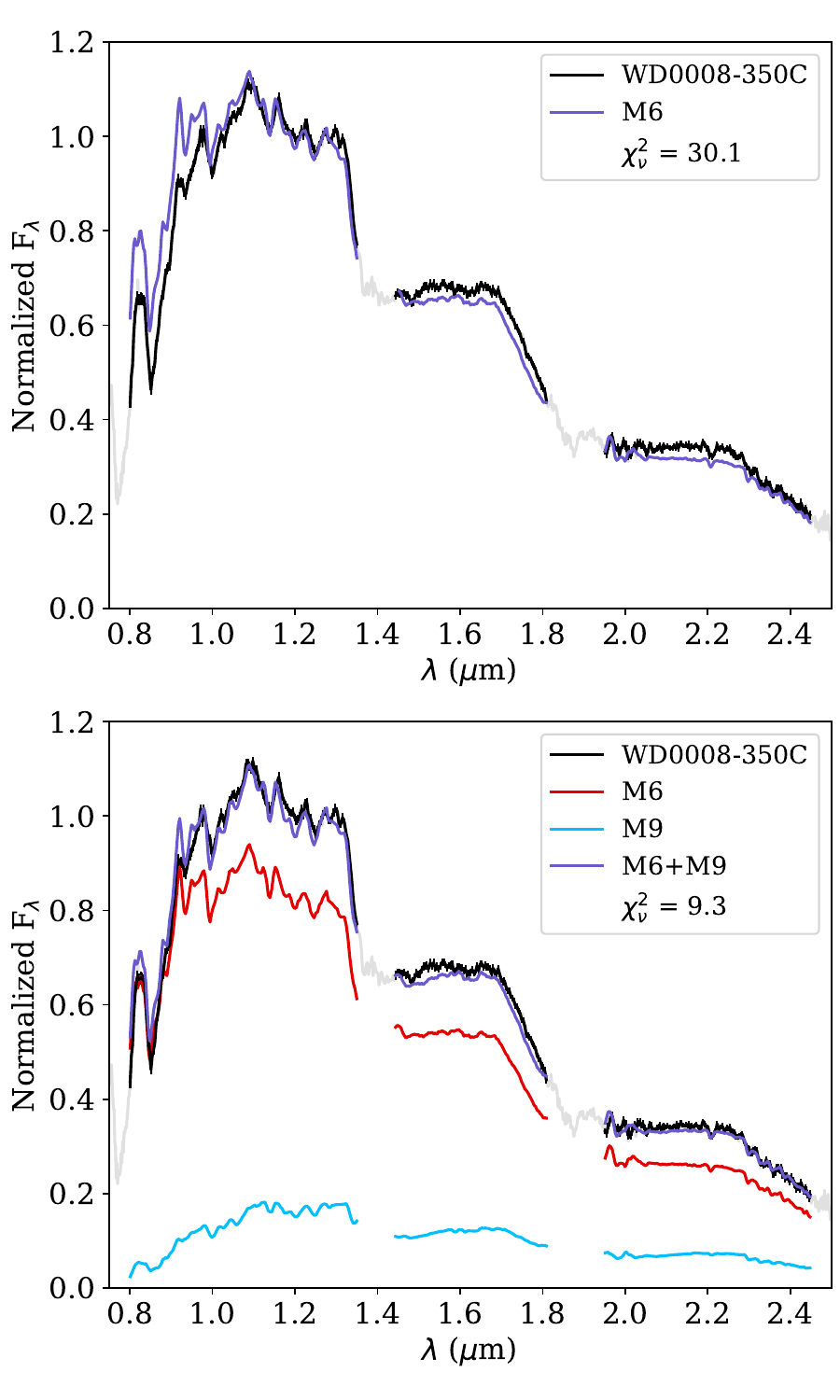}
    \caption{Single and binary template fits for the C component. The full spectrum of WD 0008-350C is shown in grey, while the regions used for the template fits are shown in black. The single fit (top) uses the M6 standard LHS 1375. The binary fit (bottom) for the C component M6+M9 uses the M9 standard LHS 2924.}
    \label{fig:companion_spectra single M6 fit and binary M6 + M9 fit}
  \end{figure}

Fig. \ref{fig:companion_spectra single M6 fit and binary M6 + M9 fit} shows the results of fitting binary templates to the C component, where it can be seen that the shorter wavelength part of the spectra is a much better fit by the binary template. We note an important caveat that the observed discrepancy is entirely attributable to the M6 template being excessively blue. Although the M6 template indeed appears too blue in certain spectral regions, it is too red at shorter wavelengths ($\lambda < 0.9\,\mu\text{m}$). This indicates that a simple instrumental effect is an unlikely sole explanation, unless such an effect can differentially impact distinct portions of the spectrum in different ways.

Using the spectral type versus mass/radius trends presented in Table 5 of \cite{2013ApJS..208....9P}, we estimate the mass of WD 0008-350C to be 0.10 M$_{\odot}$,  consistent with an M6 spectral type. However, if we adopt the mass of the outer companion as M6 + M9 then we determine a value of 0.18 M$_{\odot}$. 

We detected no variability in the TESS light curve of WD 0008-350C that could be due to a reflection effect or variability. This is likely due to the object's faintness, with a Gaia $G$-band magnitude of 18.32 which is below the effective photometric precision limit of the TESS-Gaia Light Curve (TGLC) dataset.  

In Table \ref{table1}, the RUWE for WD 0008-350C is 1.205; a threshold of RUWE $\leq$ 1.4 indicates a well-behaved solution \citep {Pearce2022RUWE}. 
The Gaia IPD parameter (the fraction of multi-peak Image Parameter Determination) transits is 0. These parameters also constrain any object that is unresolved in the Gaia astrometric data. It is consistent with the reported absence of any companions beyond an angular separation of 0.3 arcseconds. Although the template fitting procedure favours a binary solution, constraints imposed by the Gaia data limit the range of possible binary configurations. Specifically, the RUWE and IDP values from Gaia effectively rule out many potential orbital solutions. Therefore, additional high-resolution observations are necessary to confirm or refute the inferred binary nature of this source.

We also investigated the near-infrared colors of the C component of this system to see whether or not they might provide additional evidence of its multiplicity. We find $J-H$ and $H-K$ colors of 0.67$\pm$0.04 mag and 0.36$\pm$0.04 mag, respectively compared to the median $J-H$ ($0.60^{+0.06}_{-0.05}$ mag) and $H-K$ ($0.33\pm$0.05 mag) colors of $M6-M6.9$ dwarfs from \cite{2018ApJS..234....1B}, the values for WD 0008-350C are redder, but within $\sim$1$\sigma$ of the median values in both cases. While these redder generally support the binary hypothesis for this component of this system, additional observations are likely needed to determine its potential multiplicity definitively. 

\section{System Properties}

\subsection{Co-Moving Probability}
Initially, we utilised the \texttt{WiseView} tool \citep{Wiseview} to look for any additional companions which were not detected by $Gaia$ (e.g. brown dwarfs). \texttt{WiseView} blinks time-resolved \textit{WISE} W1 \& W2 coadds spanning $\sim$12 yrs to create a time-lapse video for a specified region of sky, allowing for convenient visualisation of proper motion. It also contains a \textit{Gaia} DR3 overlay, which displays the \textit{Gaia} proper motion, which allowed for easy visual comparison. Using this technique, we have ruled-out any additional low-mass companions in the system out to a separation of approximately 100\arcsec. 

Faint companions are hard to distinguish from much brighter primaries in WISE because of the low angular resolution.

 \texttt{CoMover} \citep{CoMover} was employed to determine the probability that the host white dwarf and the stellar companions are co-moving and therefore likely to be gravitationally bound. \texttt{CoMover}  uses a framework similar to the BANYAN~$\Sigma$ Bayesian classification algorithm. The co-moving probability is determined by comparing a model of the potential companion's motion against models of both the field stars and the host star. 
 
 The main difference between BANYAN~$\Sigma$  and \texttt{CoMover} is that \texttt{CoMover} only includes models of field stars (see \citealt{Gagné_2018} for more detail). 
 The \texttt{CoMover} algorithm marginalizes over any missing parameters (such as the radial velocity or parallax of the companion) to compute the probability density in $XYZUVW$ space directly, using the analytical solutions to the marginalization integrals provided in \cite{Gagné_2018}. The resulting probability densities are compared using Bayes' theorem to determine the co-moving probability.

For this system, both the inner and outer pairs exhibit a co-moving probability exceeding 99.9\%. The log probability of the field hypothesis is -11.8 for the outer companion-primary pair. This corresponds to an extremely small probability of non-association: $7.45 \times 10^{-6}$
for the outer pair. A potential limitation is that the calculated co-moving probabilities are model-dependent.  This introduces a caveat: inaccuracies in the models of either the host star or nearby field stars could lead to an overestimation of the co-moving probability.
 
\subsection{Orbital Characteristics}
We applied Kepler's laws to determine the orbital characteristics of the components within the triple system WD 0008-350ABC. Assuming circular orbits, and incorporating our stellar masses of 0.63 $\pm$0.03 M$_{\odot}$ for WD 0008-350A, 0.08 M$_{\odot}$ for WD 0008-350B, and 0.10 M$_{\odot}$ for WD 0008-350C (M6, single) and 0.18 M$_{\odot}$ (M6 + M9 binary). Using parallax and co-ordinates from the $Gaia$ mission, the companions were found to be $\sim$100 au and $\sim$3,600 au from the primary, WD 0008-350 A.

The inner companion (WD 0008-350B) was found to have an orbital period of $\sim$1,200 years, while the outer companion (WD 0008-350C) has a period of $\sim$240,000 years. We utilised the mass of the outer companion, as determined by the M6 fit, to calculate the system's periods. Although the M6 + M9 template is a superior overall fit to the spectrum, the singular M6 model was adopted here pending future confirmation whether the outer companion is single or binary in nature.

Gaia satellite measurements of the relative motion between components in a hierarchical system can be used to validate or constrain orbital estimates. 


The kinematic data in Table 1 show the proper motion of Component C to be consistent with the average proper motion of Components A and B. This suggests the motion of Component C represents the true systemic proper motion ($\mu_{\text{sys}}$) of the system.  

This supports the hypothesis that the measured offsets between Components A and B are attributable to orbital motion within the system. The measured PM difference between A and B is 5.8 mas/yr. It is comparable to the estimated relative motion of 4.5 mas/yr in a circular face-on orbit. The Gaia data alone are not sufficient to define the inclination and eccentricity of the AB orbit.

The current astrometric precision is insufficient to rule out a face-on circular orbit configuration. Distinguishing this specific orbital geometry from the systemic motion would necessitate an increase in precision to approximately $0.01 \text{ mas/yr}$.

\subsection{Long-Term Stability}




In a sample of 392 low-mass hierarchical triple stellar systems within 100 pc resolved by Gaia, the median projected separations of the inner and outer pairs were 151 and 2569 au, respectively \citep{Tokovinin_2022}. The system exhibits an 'average' separation, characterized by inner and outer pair distances of approximately $\sim$100 au and $\sim$3,600 au, respectively. 


For a hierarchical triple system with an inner separation of $a_{\text{in}}$ and an outer separation of $a_{\text{out}}$ - the point at which stability breaks down generally depends on the separation ratio $\alpha\equiv a_{\text{in}} / a_{\text{out}}$ \citep{Tory_Grishin_Mandel_2022}. $\alpha$ $\approx1$ suggests a highly chaotic system. For this system, $\alpha$ = 0.028. This system is therefore very stable, far from the threshold of instability.



\section{Conclusion}
New spectroscopic data have been presented for all components of this system. Subsequently, the WD mass is revised from M = 0.18$\pm$0.06 M$_{\odot}$ to M = 0.63$\pm$0.04 M$_{\odot}$. Our analysis yielded mass estimates of 0.08 M$_{\odot}$ and 0.10 M$_{\odot}$ (M6, single) or 0.18 M$_{\odot}$ (M6 + M9, binary) for the two outermost companions.

Our spectrally inferred masses are, respectively, slightly lower for the outer two companions when compared with the photometrically derived masses of 0.13 M$_{\odot}$ and 0.14 M$_{\odot}$ \cite{Tokovinin_2022} and 0.11 M$_{\odot}$ and 0.14 M$_{\odot}$ \cite{Kervella}. 

We determine the following:
a total age of $4.67^{+3.58}_{-0.85}$ Gyr   for the system; a Main Sequence age of $2.15 _{-1.04}^{+3.83}$ Gyr for the progenitor; an initial stellar mass of the WD0008-350A  estimated at $1.65 _{-0.44}^{+0.56}$M$_{\odot}$.

The inner companion's spectral class is determined to be M8. A spectral binary template of M6+M9 best matches the outer companion's spectral type. The template fitting procedure  favours a binary stellar configuration. However, the range of viable binary solutions is significantly constrained by the Gaia data. Further analysis, such as high-resolution imaging or radial velocity monitoring, may confirm the binary or single nature of this object.  The TESS light curve of WD 0008-350C showed no detectable photometric variability. 

This stellar configuration represents a relatively rare hierarchical system, comprising either three or four components. Such systems are invaluable for estimating the age of their ultracool stellar companions, as their ages can be directly linked to the more accurately determined age of the white dwarf primary.
Should this system be definitively confirmed as a WD + 3 hierarchical configuration, its occurrence rate would fall below the 1.9\% for quadruple systems as reported by  \cite {10.1111/j.1365-2966.2008.13596.x}. This highlights its unique nature and importance for refining our understanding of multiple-stellar system demographics.

With $\alpha$ =0.029, this system represents a stable configuration.

Additional observational data beyond those provided by Gaia are required to constrain the orbital inclinations and eccentricities in this system.

The current astrometric precision is insufficient to exclude a face-on circular orbit configuration.

\section{Acknowledgements}
The Backyard Worlds: Planet 9 team would like to thank the many Zooniverse volunteers who have participated in this project. We would also like to thank the Zooniverse web development team for their work creating and maintaining the Zooniverse platform and the Project Builder tools. This research was supported by NASA grant 2017-ADAP17-0067. This material is supported by the National Science Foundation under grant No. 2007068, 2009136, and 2009177.
Based on observations collected at the European Organisation for Astronomical Research in the Southern Hemisphere under ESO programme(s) 113.26HT.001. Data were also taken by Visiting Astronomers at the Infrared Telescope Facility, which is operated by the University of Hawaii under contract 80HQTR24DA010 with the National Aeronautics and Space Administration.
This work has made use of data from the European Space Agency (ESA) mission $Gaia$ (https://www.cosmos.esa.int/gaia), processed by the Gaia Data Processing and Analysis Consortium (DPAC, https://www.cosmos.esa.int/web/gaia/dpac/consortium). Funding for the DPAC has been provided by national institutions, in particular the institutions participating in the $Gaia$ Multilateral Agreement, This publication makes use of data products from the {\it Wide-field Infrared Survey Explorer}, ({\it WISE}) which is a joint project of the University of California, Los Angeles, and the Jet Propulsion Laboratory/ California Institute of Technology, funded by the National Aeronautics and Space Administration. RMO is funded by INTA through grant PRE-OBSERVATORIO and acknowledges support from project PID2023-146210NB-I00 funded by MICIU/AEI/10.13039/501100011033 and by ERDF/EU.

\paragraph{Software:}
CalcTool {(https://www.calctool.org/astrophysics)};
CoMover \citep{CoMover};
Lightkurve \citep{2018ascl.soft12013L}
WiseView \citep{Wiseview}; \textsc{wdwarfdate} \citep{Kiman_2022}

\bibliography{references.bib}
\bibliographystyle{aasjournal}

\end{document}